\documentstyle[aas2pp4,twoside,aabib]{article}
\input epsf

\def\s2n{S^{\prime}/N}

\newcommand{\Grad}{\bf{\nabla}}

\begin{document}
\title{Turbulent Fragmentation and the Initial Conditions for Star Formation}
\author{Paolo Padoan}
\affil{Harvard-Smithsonian Center for Astrophysics, Cambridge, MA 02138, USA}
\author{\AA ke Nordlund}
\affil{Copenhagen Astronomical Observatory, and Theoretical Astrophysics Center, 
2100 Copenhagen, Denmark}
\author{\"{O}rn\'{o}lfur Einar R\"{o}gnvaldsson}
\affil{Nordic Institute for Theoretical Physics, 2100 Copenhagen, Denmark }
\author{Alyssa Goodman}
\affil{Harvard-Smithsonian Center for Astrophysics, Cambridge, MA 02138, USA}

\begin{abstract}
Super--sonic turbulence fragments molecular clouds (MC) into a very complex
density field with density contrasts of several orders of magnitude.
A fraction of the gas is locked into dense and gravitationally bound cores,
which collapse as proto--stars. This process can be studied with numerical
simulations of super--sonic self--gravitating turbulence. 

In this work, we use numerical simulations of magneto--hydrodynamic (MHD),
super--sonic, super--Alfv\'{e}nic and self--gravitating turbulence to
compute the mass distribution of collapsing proto--stellar cores, which 
are selected as local density maxima. We find that the mass distribution
of collapsing cores is consistent with the stellar initial mass function
(IMF), suggesting that super--sonic turbulence may be responsible for the 
generation of the IMF.

To support this conclusion we also show that the physical properties
of the numerically selected cores are in agreement with the properties
of observed NH$_3$ cores and that their magnetic field strength is
consistent with Zeeman splitting measurements.

In turbulent MCs, star formation occurs via the gravitational collapse
of super--critical cores, formed by the turbulent flow, sub--critical
cores being irrelevant for the process of star formation.

\end{abstract}

\keywords{
turbulence -- ISM: kinematics and dynamics; radio astronomy: interstellar: lines
}

\section{Introduction}

Despite early observational evidence of the highly 
super--sonic nature of molecular cloud (MC) turbulence (Zuckerman \& Palmer 1974), 
there have been attempts to simplify the problem of star formation by: 
i) assuming that super--sonic turbulence behaves like sub--sonic laboratory 
turbulence; or by ii) focusing on the process of the formation of one single 
star, neglecting its larger scale environment. 
The first type of simplification has allowed theoreticians
to consider super--sonic turbulence as a source of internal pressure,
as suggested by Chandrasekhar (1958) for sub--sonic turbulence (e.g. 
Bonazzola et al. 1987, 1992). This simplification puts aside 
the most important characteristic of super--sonic turbulence, that is,  
the generation of a very fragmented density field by a complex 
system of turbulent shocks. The second type of simplification has allowed for 
the generation of analytical and numerical models for the 
collapse of proto--stars, based on quasi statically evolving proto--stellar 
cores (e.g. Shu, Adams \& Lizano 1987), which are hard to reconcile with the
turbulent medium from which they are formed.
A complementary way to model the dynamics of MCs avoiding the problem of 
super--sonic turbulence has been the assumption that the
observed line width of molecular transitions is not due to turbulence, but
to Alfv\'{e}n waves, that is, oscillations of a rather strong magnetic field
(Arons \& Max 1975; Zweibel \& Josafatsson 1983; Elmegreen 1985; Falgarone \&
Puget 1986). However, even in this scenario, random motions and 
super--sonic shocks are generated by super--sonic compressions along 
magnetic field lines, and so the process of turbulent fragmentation is unavoidable, 
irrespective of the magnetic field strength.

Recent numerical studies of super--sonic magneto--hydrodynamic 
(MHD) turbulence have shed new light on the physics of turbulence. 
The most important results are:
i) Super--sonic turbulence decays in approximately one dynamical 
time, independent of the magnetic field strength (Padoan \& Nordlund 1997, 1999;
MacLow et al. 1998; Stone, Ostriker \& Gammie 1998; MacLow 1999);
ii) The probability distribution function of gas density in isothermal turbulence is well 
approximated by a Log--Normal distribution, whose standard deviation 
is a function of the rms Mach number of the flow (Vazquez--Semadeni 1994;
Padoan 1995; Padoan, Jones \& Nordlund 1997; Scalo et al. 1998; 
Nordlund \& Padoan 1999; Ostriker, Gammie \& Stone 1999);
iii) Super--sonic isothermal turbulence generates a complex system 
of shocks that fragment the gas very efficiently into high density
sheets, filaments, and cores; 
iv) Super--Alfv\'{e}nic turbulence provides a good description of the
dynamics of MCs, and an explanation for the origin of dense cores with
magnetic field strength consistent with Zeeman splitting measurements. 
(Padoan \& Nordlund 1997, 1999). 

We call ``turbulent fragmentation'' the process leading to the generation 
of high density structures (and low density ``voids'') by turbulent shocks. 
In previous works we investigated the possibility that MCs are fragmented 
primarily by random super--sonic motions, assuming that self--gravity is important
only in the densest regions, so we used numerical simulations
of turbulent flows where the effects of self--gravity were intentionally
disregarded (Padoan, Jones \& Nordlund 1997; Padoan et al. 1998, 1999; 
Padoan \& Nordlund 1997, 1999; Padoan, Zweibel \& Nordlund 2000; Padoan et al. 2000). 
We found that many observational properties of MCs are indeed consistent with
the properties of turbulent flows, which supports the suggestion that gravity
is important only in the densest regions formed by turbulent shocks.

In this work we include the description of self--gravity, in simulations 
of three dimensional super--sonic MHD turbulent flows (\S~2). Self--gravity
is here required because we are interested in selecting gravitationally bound 
(collapsing) protostellar cores, formed by the process of turbulent
fragmentation. The mass
distribution of proto--stellar cores, selected numerically as local
maxima in the gas density, is computed in \S~3. The value
of this result is then confirmed by the comparison between the 
numerically selected cores and the observed NH$_3$ cores, in \S~4. 
The simulation also allows us to predict core properties, which 
are difficult to measure observationally and are therefore not 
well known, such as the magnetic field strength (\S~5). 
In \S~6 we discuss the results of this work, and in
\S~7 we list our main conclusions.

\section{The Numerical Simulations}

\subsection{The Equations}

We solve the compressible MHD equations:

\def\vv{{\bf v}}
\def\jj{{\bf j}}
\def\bb{{\bf B}}
\def\lnr{\ln\rho}
\def\div{\nabla\cdot}

\begin{equation}
\label{0}
{\partial \ln\rho \over \partial t} + \vv \cdot \nabla\lnr = - \div \vv,
\end{equation}

  \begin{equation}
   {\partial{\vv} \over \partial t}
   + {\vv\cdot\nabla\vv}
  =
   - {P\over\rho} \nabla \ln P
   + {1\over\rho} {\jj} \times {\bb} - {\Grad}{{\Phi}}
   + {\bf f},
  \label{1}
  \end{equation}

\begin{equation}
\label{4}
{\partial e \over \partial t} + {\vv} \cdot \nabla e = - {P \over \rho} \div 
{\vv} + Q_{\rm dissipation} + Q_{\rm radiation},
\end{equation}

\begin{equation}
{\partial{\bb} \over \partial t} = \nabla\times\vv\times\bb,
\label{2}
\end{equation}

\begin{equation}
\jj = \nabla\times\bb,
\label{3}
\end{equation}

\begin{equation}
\nabla^2 \Phi = C \rho
\end{equation}

\noindent
plus numerical diffusion terms, and
with periodic boundary conditions. $\vv$ is the velocity, $\bb$ the
magnetic field, $\Phi$ the gravitational potential, ${\bf f}$ an external 
random force, $e$ the internal energy, $P = \rho T$ is the 
pressure at $T \approx$ const. The isothermal approximation (see discussion
in Padoan, Zweibel \& Nordlund 2000) makes the energy equation (3)
redundant, and so the $Q_{\rm dissipation}$ and $Q_{\rm radiation}$ 
are not defined here. The constant $C$ is given by:

\begin{equation}
C = \frac{4 \pi G l_0^2 \rho_0}{v_0^2},
\end{equation}
where the velocity is measured in units of $v_0$, length in units of $l_0$, 
time in units of $l_0/v_0$, and density in units of $\rho_0$.
If the velocity is measured in units of the sound speed $c_s$,
and the computational box size is $L_0=Nl_0$, where $N$ is the number
of computational cells along one dimension of the numerical mesh,
the constant $C$ can be expressed as:

\begin{equation}
C = \frac{1}{N^2}\left(v_{ff}^2 \over{ c_{s}^2} \right),
\end{equation}
where $v_{ff}$ is the free--fall time.

\subsection{The Model}

For the purpose of this paper we have run a numerical simulation
of super--sonic, super--Alfv\'{e}nic and self--gravitating MHD 
turbulence, by solving numerically the equations given above.
As in our previous works, the initial density and magnetic fields
are uniform; the initial velocity is random, generated in
Fourier space with power only on the large scale. We also apply
an external random force, to drive the turbulence at a roughly constant
rms Mach number of the flow. This force is generated in Fourier
space, with power only on small wave
numbers ($1<k<2$), as the initial velocity.

We let the flow evolve for one dynamical time. In our previous 
simulations without self--gravity, we usually let the flow evolve 
and relax for many dynamical times, since all density
structures are transient. With self--gravity this
is not possible because the flow does not statistically relax,
and continues to generate an increasing number of collapsing
cores, and to accrete mass around them. Since our numerical
resolution (128$^3$ numerical mesh) allows only the description
of the initial phase of the collapse of single cores (\S~2.3), results
are progressively inaccurate at later times, when the numerical
resolution cannot cope with the exceedingly high density. We therefore 
interrupt the simulation at a time when most cores are just
recently formed and start to collapse, which is about one dynamical
time of the large scale, $t_{dyn}=L_0/\sigma_v$, where $L_0$
is the linear size of the computational box, and $\sigma_v$ is
the rms flow velocity. 

Periodic boundary conditions and large scale external forcing
are justified by the fact that we simulate a region
of turbulent flow inside a larger turbulent molecular 
cloud. The rms Mach number of the flow is 
${\cal M}_s\approx 10$, which corresponds to a linear size 
$L_0\approx 5$~pc, and an average gas density $<n>\approx 900$~cm$^{-3}$, 
using empirical Larson type relations (e.g. Larson 1981; Myers 1983; Fuller 
\& Myers 1992). The average magnetic field in this model is rather weak, as
justified by our previous work (Padoan \& Nordlund 1997, 1999), and 
such that the average magnetic energy is of the order of the
average thermal energy. Assuming a kinetic temperature $T=10$~K,
the rms flow velocity is $\sigma_v\approx 2.5$~km/s, and the
average magnetic field strength is $\langle B\rangle \approx 4.5$~$\mu$G.
Despite this low value of $\langle B\rangle$, strongly magnetized cores are
formed by the process of turbulent fragmentation (with field strength 
sometimes in excess of 100~$\mu$G), due to local compressions in the 
turbulent flow.

\subsection{Proto--stellar Cores in Numerical Simulations of Super--Sonic Turbulence}

Star forming cores in MCs have typical gas density of about 10$^4$~cm$^{-3}$,
and are part of larger clouds with density of the order of  10$^3$~cm$^{-3}$.
Although MCs have internal structure that spans a continuous range of gas 
density, these values are representative of clouds and cores on the scale
probed by $^{13}$CO and NH$_3$ transitions respectively. 
The initial density of individual proto--stellar cores
(approximately 10$^5$~cm$^{-3}$) is close to the maximum density achieved
by a finite difference MHD code on a 128$^3$ numerical mesh; when a scale
of 5--10~pc and an average density of approximately 10$^3$~cm$^{-3}$ are 
simulated. In such a simulation, with an rms Mach number ${\cal M}_s\approx 10$,
a density contrast of about 5 orders of magnitude is typically achieved
from a minimum density of 1~cm$^{-3}$ to a maximum density of 10$^5$~cm$^{-3}$.
Higher densities, characteristic of collapsing proto-stellar cores, cannot 
be achieved with such a numerical tool. In a few years numerical simulations
of the same sort, in a larger numerical mesh with size of the order of 1000$^3$,
will become generally available and will allow the numerical description
of the initial phase of the collapse of proto--stellar cores.  
The future application of smooth particle hydrodynamic (SPH) codes 
to the study of super--sonic turbulence with strong radiative cooling
(Klessen, Heitsch \& Mac Low 2000), or the future development of MHD codes 
on an adaptive mesh (Truelove et al. 1998) are also promising.
Meanwhile, it is possible to investigate the role of self--gravitating 
turbulence in the formation of proto--stellar cores, even if only the
initial conditions for the gravitational collapse of each core can be 
described numerically. The formation of collapsing cores in numerical simulations
of super--sonic turbulence, and their rate of disruption by the turbulent
flow, has been previously discussed by Klessen, Heitsch \& Mac Low (2000), 
and Heitsch, Mac Low \& Klessen (2000). These works focus on the issue
of turbulent and magnetic support against local gravitational collapse. 
They conclude that local gravitational collapse cannot be avoided unless the
turbulence is driven on scales smaller than the local Jeans length in the 
densest regions, or the magnetic field provides magnetostatic support.

\section{The Mass Distribution of Collapsing Cores}

In order to select collapsing cores from the numerical simulations,
we have defined as cores density fluctuations with an amplitude
of at least 2.5\%, 

\begin{equation}
\frac{\rho_{max}-\rho_{iso}}{\rho_{iso}}\ge 0.025
\end{equation}
where $\rho_{max}$ is the density value at the local density maximum
that defines the fluctuation, and $\rho_{iso}$ is the density value
of the density isosurface that delimits the mass assigned to the 
fluctuation. The isosurface is determined as the smallest density such that
the whole region inside the isosurface contains only one density maximum
that exceeds the isosurface value by 2.5\%.

Density fluctuations of a 2.5\% amplitude grow to only a 50\% amplitude 
while the collapsing background has increased its density of about 3 orders 
of magnitude (Tohline 1980). Density fluctuations of smaller amplitude are 
unlikely to collapse away from the background before they (or the background)
become rotationally supported. However, this choice of the 2.5\% 
amplitude threshold is somewhat arbitrary, as discussed below (see \S~6).
According to the definition of core given above, we do not select 
cores that are fragmented into smaller ones. If at least two smaller 
cores are found inside a larger one, the latter is excluded
from the sample. Moreover, only cores with gravitational energy
in excess of the sum of thermal and magnetic energies are included
in the sample because we are interested only in collapsing
proto--stellar cores, that is, in super--critical ones. Sub--critical
cores are transient density enhancement, which will re-expand
or be destroyed by the turbulent flow, or will later accrete more
mass in a dynamical time from the random turbulent flow to become 
super--critical and collapse. There is only a very small chance 
that sub--critical cores, formed by a turbulent flow, are found 
in equilibrium and are given the opportunity to maintain such
equilibrium, for as long as the ambipolar drift
time (cf. Nakano 1998). Therefore, although sub--critical cores can exist as 
transient structures, they are almost
irrelevant to the process of star formation, contrary to the
picture once proposed in the literature (Shu, Adams \& Lizano 1987).

The mass distribution of collapsing cores, selected from the MHD
simulations (MHD cores), is plotted in Figure~1 (left panel), as the number of
cores per logarithmic mass interval. Cores above 2~$M_{\odot}$ are selected
by scaling the computational box to a physical size of 15~pc. Cores smaller
than 2~$M_{\odot}$ are selected after rescaling to a physical size of 5~pc.
In order to match the two mass distributions, the total number of cores
from the small scale has been multiplied by a factor equal to the ratio
of the total mass contained in the two models. Since we scale the physical
values of the average gas density in the computational box according to
a Larson type relation, $n\propto L^{-1}$, the 5~pc model has an average 
density 3 times larger than the 15~pc model, and thus a total mass 9 times 
smaller.
The number of cores selected from the 5~pc model has therefore been multiplied
by a factor of 9. Each mass distribution extends above and below 2~$M_{\odot}$,
and match almost exactly in the mass range where they overlap. In order to
plot them together, we have used only the mass range above 2~$M_{\odot}$
for the 15~pc model and the mass range below 2~$M_{\odot}$ for the 5~pc model. 

The distribution in Figure~1 is a power 
law between 2 and 200 $M_{\odot}$, with a slope, $\Gamma=1.34$, 
consistent with the Salpeter stellar initial mass function (IMF), 
$\Gamma=1.35$. At about 2~$M_{\odot}$ the mass distribution
flattens, and turns around at about 0.5~$M_{\odot}$ (in these 
logarithmic units). The scale--free behavior at large and intermediate
masses is due to the process of turbulent fragmentation. The 
initial flattening at about 2~$M_{\odot}$ is caused by the 
magnetic pressure: for increasingly smaller masses,
an increasing number of small density enhancements assembled by 
turbulence are found to be sub--critical (magnetic plus thermal energies 
in excess of the gravitational energy), and are therefore 
excluded from the sample because they are not collapsing. The 
cutoff at the smallest masses is mainly due to the thermal pressure,
because most of the smallest cores are smaller than their own
Jeans mass and are therefore excluded from the sample of collapsing cores.

In the plot in Figure~1 (left panel) there is no core below 
0.2~$M_{\odot}$. This particular value is set by the numerical resolution
of the simulation that does not allow one to discern anything smaller. 
The actual cutoff, and also the mass of the turn--around of
the distribution, should be somewhat smaller. Moreover, it is very likely 
that even individual proto--stellar cores fragment at least into a binary
system. If the mass distribution of each member of a binary system is 
plotted, instead of the mass distribution of proto--stellar cores,
the low mass end of the distribution is much more populated. As an 
example, we have computed the mass distribution of single components
of the binary systems, assuming that each core fragments into two
components, whose mass ratio is a uniformly distributed random
variable with values between 0 and 1. The result is plotted on
the right panel of Figure~1. The mass distribution of single
components has its maximum value at approximately 0.3~$M_{\odot}$, 
it is almost symmetric around the maximum, and it extends down to
sub--stellar masses with a probability comparable to the one of the
most massive stars. Expressed in linear mass units (dotted line), 
the mass distribution is almost flat below the maximum, which is
found at approximately 0.2~$M_{\odot}$.

Finally, the whole distribution should be shifted somewhat towards smaller
masses, if it is to be compared with the stellar IMF (for example 
Luhman et al. 2000) since proto--stellar collapse is always accompanied 
by mass loss.

\section{Numerical Cores and Observed NH$_3$ Cores}

In the previous section we have found that the mass distribution
of collapsing cores in the MHD simulation (`MHD cores') is consistent
with the stellar IMF. In order to confirm the validity of
this result, it is useful to compare the physical properties
of the MHD cores, with the properties of cores selected observationally
via molecular line transitions. 
We use the NH$_3$ core sample from Jijina, Myers \& Adams (1999),
which is a literature compilation of 264 NH$_3$ cores. For the purpose 
of this work, we use 149 cores not associated with stellar clusters,
as defined in Jijina, Myers \& Adams' sample, because they offer a more 
appropriate description of the initial conditions of star formation.

Figures~2 and 3 show the probability distribution
of the radius, non--thermal line width, velocity gradient
(rotation), and gas density of both NH$_3$ and MHD cores 
(notice that all the histograms are in log units).
Numerical and observed cores are very similar in size,
amount of turbulence, rotation, and density. NH$_3$ cores
have slightly larger line width, which could be due to velocity
superposition along the line--of--sight of physically unrelated gas.
Average values and variances of each histogram 
are listed in Table~1.
Slightly differences between the properties of MHD and NH$_3$
cores are expected, because the properties of the MHD cores 
are here obtained directly from the three dimensional numerical 
data--cubes of density and velocity field. We have not tried
to reproduce exactly the observational procedure by solving the
radiative transfer through the data--cube. This more detailed 
comparison with the observations, via the computation of
synthetic spectral maps, has been the subject of other works
(Padoan et al. 1998, 1999, 2000). 

In order to further compare the two core samples, we consider
the ``Type 2'' correlation between size and line width (single--tracer
multiple--cloud --Goodman et al. 1998), which we plot in
Figure~4. The NH$_3$ cores without a stellar cluster 
association are represented by open squares, and the MHD cores by 
asterisks. A least square fit is also shown for both samples. The 
line width versus size relations are remarkably similar; the slope 
is $(0.56\pm0.22)$ for the NH$_3$ cores, and $(0.57\pm0.15)$ for
the MHD cores (Table~1).

\section{Magnetic Fields in Proto--Stellar Cores}

The magnetic field strength, $B$, in proto--stellar cores is usually
unknown, because it is very difficult to measure observationally.
In order to compare $B$ in MHD cores with the observations, we need 
to refer to the few available detections and upper limits of $B$, 
which are based on OH or CN Zeeman splitting
measurements, although such measurements apply to a variety of
scales, and not just to proto--stellar cores. 
In Figure~5 we have plotted B
versus the estimated H$_2$ column density, using a sample
of Zeeman splitting measurements from Bourke et al. (2000), 
which contains the previous sample by
Crutcher (1999), a number of original detections and upper limits
and the recent detections by Crutcher \& Troland (2000) in L1544 and
by Sarma et al. (2000) in NGC6334. Asterisks represent
detections and triangles upper limits (19 detections and 31 upper limits).
From the point of view of this work (see below), a conservative choice
is to assume that the magnetic field strength is in all cases very
close to the estimated upper limits. The dashed line in Figure~5
is a least square fit to the observations, where upper limits have 
been treated as if they were detections. 

The numerical cores, represented by squares in Figure~5, 
follow almost exactly the same $B$--$N_{col}$ relation as the observations
(the solid line is their least square fit). We have also plotted
in the same figure the line that marks the equality between
magnetic and gravitational energies, ${\cal M}={\cal W}$.
All magnetic field detections but one are below that line, which means that
all regions observed have gravitational energy in excess of the
magnetic energy, and only 3 of the 31 upper limits and 1 of the 19 detections
are above the line.
Of the 80 numerical cores, only 3 are above the line. All
numerical cores, including those three, are super--critical (gravitational
energy in excess of the magnetic energy) and have been selected as such.
They can be occasionally found above the line because they are elongated 
along the direction of the magnetic field (aspect ratio 2--3), which 
decreases their critical mass, relative to that of a spherical core 
with the same $B$ (McKee et al. 1993).
Notice that several MHD cores with relatively large values of B
are found, even in excess of 100~$\mu$G, although the average
$B$ in the simulation is only 4.5~$\mu$G, making
the large scale turbulence super--Alfv\'{e}nic
(Padoan \& Nordlund 1999).

\section{Discussion}

In this work we have found that the mass distribution of collapsing 
cores is practically indistinguishable from the stellar IMF.
This suggests that turbulent fragmentation is an
essential ingredient in the generation of the stellar IMF.
The computation of the mass distribution of cores relies
on the selection of local maxima of the gas density in 
numerical simulations of super--sonic turbulence. 
This method contains some uncertainties,
the most significant being the assumption
that any single core, corresponding to a density fluctuation
beyond a threshold amplitude, collapses as a single proto--star
or as a binary system,
if it is not initially fragmented into smaller fluctuations
with amplitude larger than the same threshold. The particular
value of the threshold of 2.5\% is justified by the fact that fluctuations
of smaller amplitude would hardly grow before they become 
rotationally supported. However, the true value could be
anything between 1\% and 4\%, and may also have spatial 
variations. A smaller amplitude threshold produces a steeper
mass distribution than a larger threshold, because it selects
a larger number of smaller cores. The slope, which we found
to be 1.34 for a 2.5\% threshold, could therefore be anything
between 1.2 and 1.5. 

Recent observational data seem to confirm the present
result, that is, turbulent fragmentation generates a mass 
distribution of proto--stellar cores very similar to the 
stellar IMF. Dust continuum observations (Motte, Andre \& 
Neri 1998) and high density molecular tracers (Ohnishi et al. 2000), 
have shown that dense cores in MCs have a mass distribution 
consistent with the stellar IMF. 

We have considered only super--critical cores,
because they collapse and form stars. This is in sharp
contrast with the usual description of low mass star
formation as the result of the quasi--static evolution
of sub--critical cores. The primary reason we do not 
consider sub--critical cores is that they are transient 
structures in a super--sonic turbulent flow. Sub--critical 
cores disperse, re--expand, or accrete more mass
from the turbulent flow and become super--critical and
collapse. The last possibility --accreting more mass-- is 
in fact the usual mode
of formation of super--critical cores, and it is a 
dynamical process, controlled by the turbulent flow.
It is highly unlikely that a sub--critical core is
formed in equilibrium at all, and even more unlikely that 
it is left undisturbed for as long as the ambipolar
drift time--scale, as assumed in many low mass star
formation theories.

We have shown that a super--sonic turbulent flow
generates super--critical cores of very small mass,
in the same proportion as required by the stellar IMF.
There is, therefore, no ``need'' to form low mass stars from 
sub--critical cores. This new point of view on star
formation suggests that star formation is a fast process,
as also proposed by Ballesteros--Paredes, Hartmann \& 
V\'{a}zquez--Semadeni (1999), Elmegreen (2000), and 
Hartmann (2000). Notice, however,
that fast does not mean efficient. As an
example, the mass distribution of collapsing cores
presented in Figure~1 contains only $\approx$2\% 
of the total mass in the simulation, which means
that in a typical MC, after one dynamical time of the
large scale, only a few percent of the total mass
is locked into collapsing cores. Star formation is
therefore inefficient on the scale of MCs because
it occurs only on a small fraction of the total mass. 
This is ultimately the most important feature of super--sonic 
turbulence: the density field fragments in a very 
intermittent way, such that a significant mass
fraction is found at densities irrelevant for the
process of star formation, or inside very small dense cores 
that are sub--critical and therefore transient (they do not
form stars).

The fast evolution of cores is also suggested by an 
extremely small fraction of starless cores in the 
sample of NH$_3$ cores by Jijina, Myers \& Adams
(1999), and by the observed chemical abundances
in dense cores (Bergin et al. 1997).
Finally, we have shown that all Zeeman splitting 
measurements are consistent with cores being
super--critical (Figure~5).

\section{Conclusions}

We have computed physical properties of collapsing
cores, in a simulation of super--sonic, super--Alfv\'{e}nic,
self--gravitating MHD turbulence. The main results are:
i) The mass distribution of collapsing MHD cores is consistent with
the stellar IMF; ii) The size, density, line width and velocity 
gradient of MHD and NH$_3$ cores are very similar; iii) MHD 
cores and observed NH$_3$ cores show the same correlation 
between line width and size; iv) Proto--stellar cores with significant 
magnetic field strength, even in excess of 100~$\mu$G, are formed in 
a flow with a low average field strength, 
$\langle B\rangle\approx 4.5$~$\mu$G; v) MHD cores show the same 
relation between magnetic field strength and column density as 
estimated with Zeeman splitting measurements.

The general conclusion is that turbulent fragmentation is essential
to understand the generation of the stellar IMF. Furthermore,
star formation, independent of the stellar mass, occurs via the
collapse of super--critical cores formed inside super--sonic
turbulent MCs, sub--critical cores being irrelevant for the process
of star formation. This point of view conflicts with many years of 
literature on star formation, but it is very promising. It has 
become accessible to investigation only in the past few 
years, because of important advances in the field of numerical 
fluid--dynamics, due to the increased power of super--computers 
and three--dimensional visualization software.



\acknowledgements

This work was supported by NSF grant AST-9721455.
\AA ke Nordlund acknowledges partial support by the Danish National 
Research Foundation through its establishment of the Theoretical 
Astrophysics Center.

\clearpage


\clearpage

\onecolumn

{\bf TABLE AND FIGURE CAPTIONS:} \\

{\bf Table~1:} Average values of core property distributions. \\  
               
\begin{figure}
\centerline{\epsfxsize=10cm \epsfbox{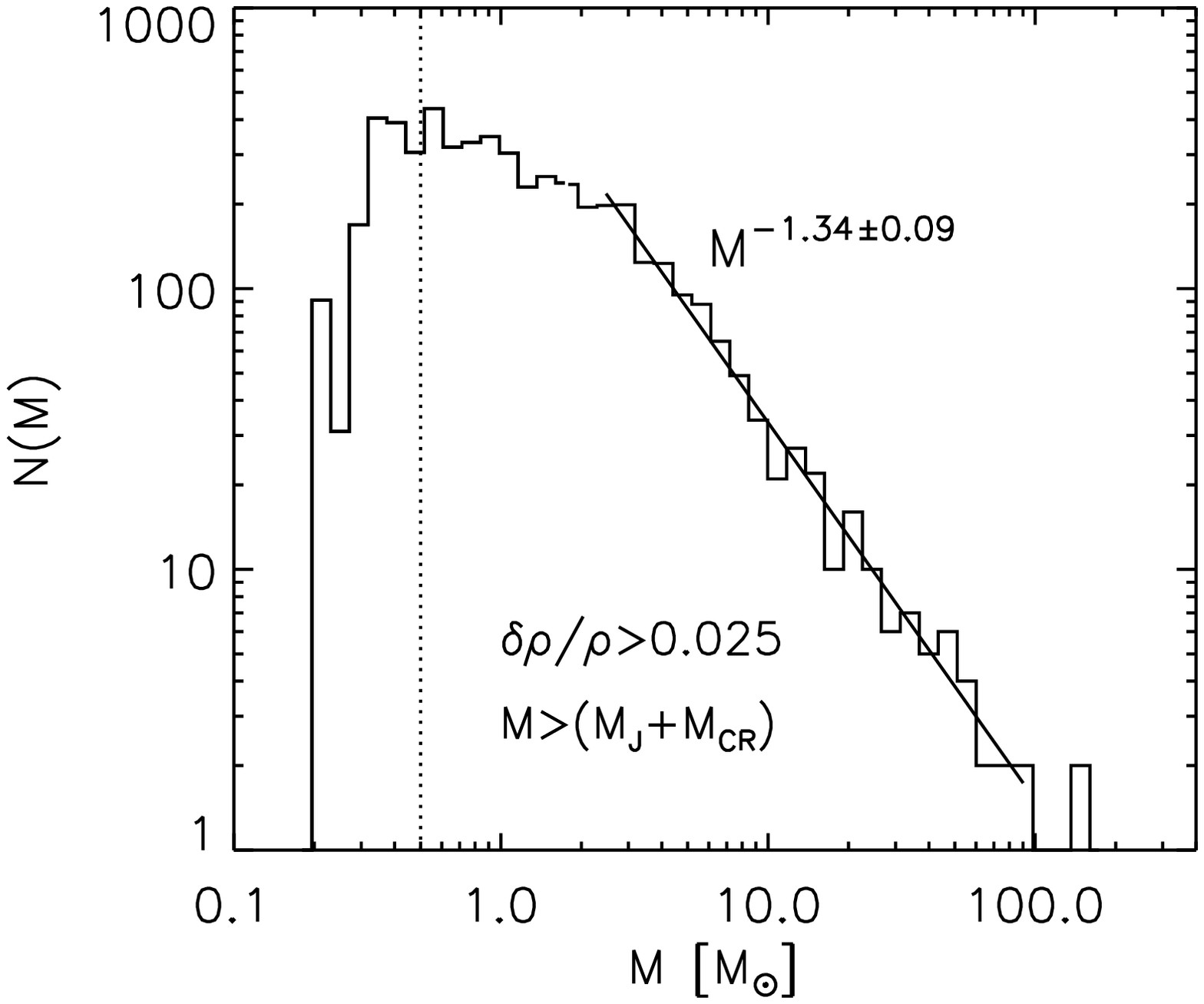}
\epsfxsize=10cm \epsfbox{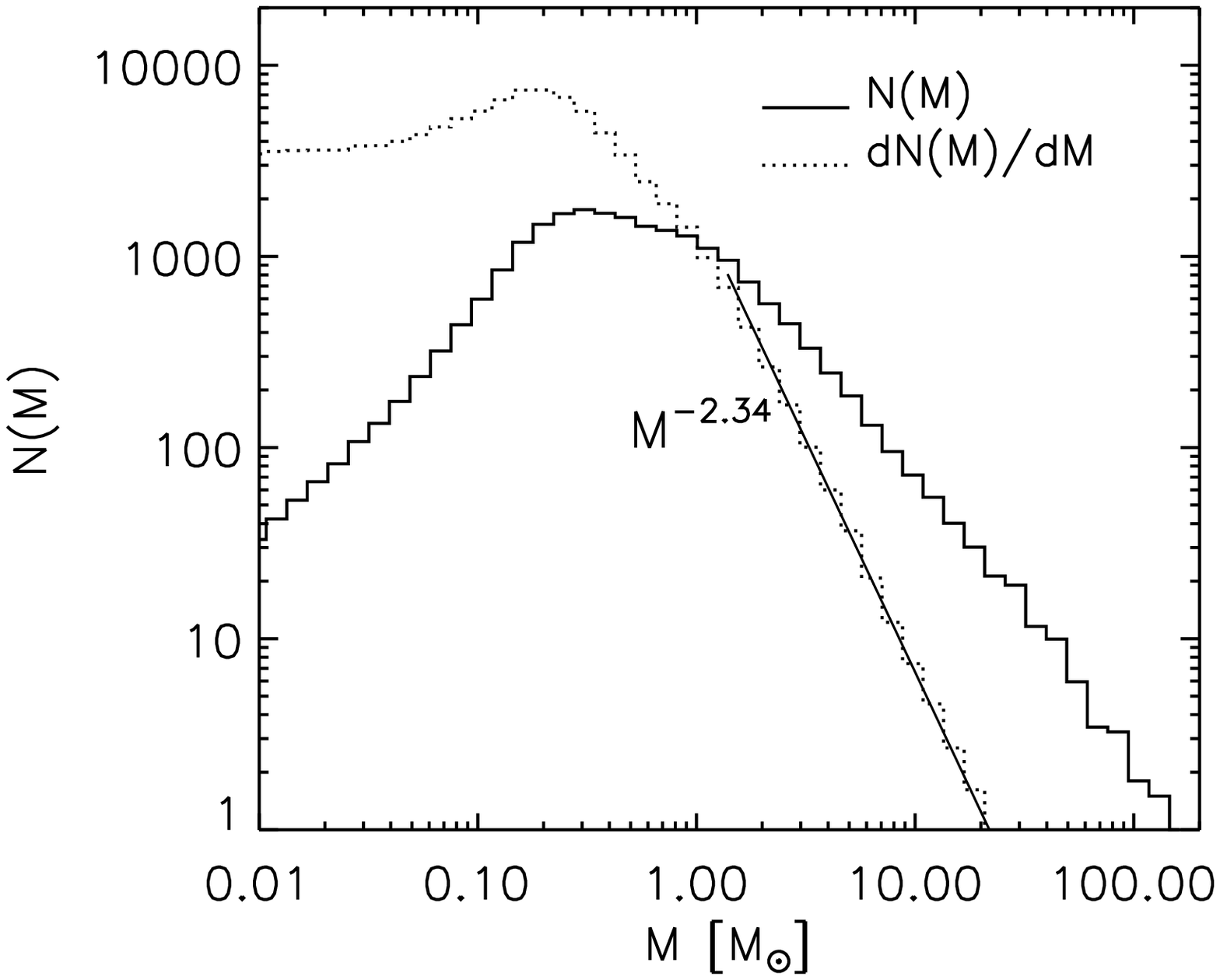}}
\caption[]{}
\label{fig1}
\end{figure}

{\bf Figure \ref{fig1}:} Left: Mass distribution of MHD cores. The 
selection of cores is incomplete at masses $<0.5M_{\odot}$ (vertical 
dotted line), and no core is found below $<0.2M_{\odot}$, because of 
the limited numerical resolution. 
Right: Mass distribution of single components of the binary systems, assuming
that all cores fragment into a binary system, and that the mass ratio of the
two components is a uniformly distributed random variable with values 
between 0 and 1. The dotted line shows the distribution computed in 
linear mass units. \\ 

\begin{figure}
\centerline{\epsfxsize=18cm \epsfbox{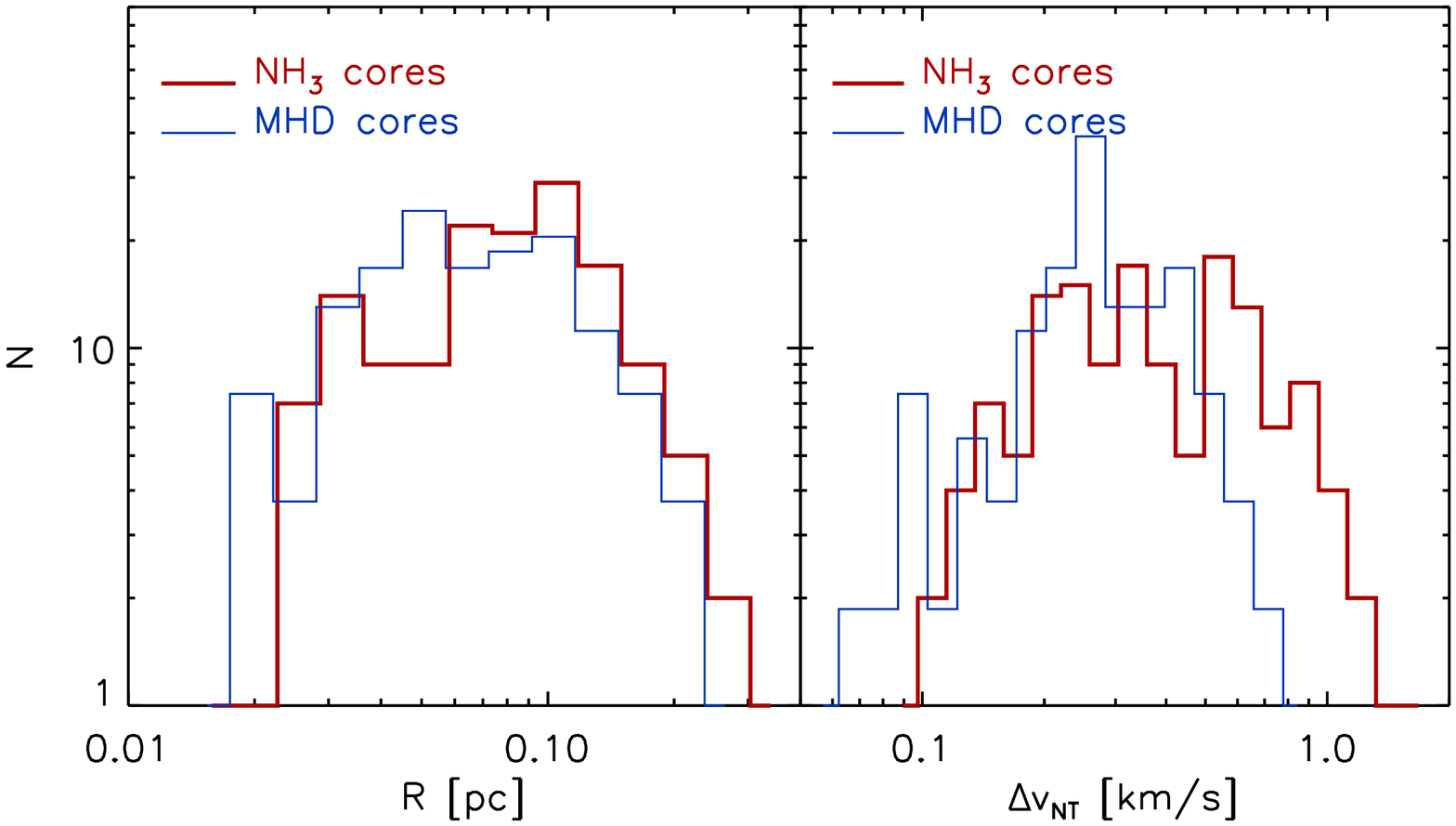}}
\caption[]{}
\label{fig2}
\end{figure}

{\bf Figure \ref{fig2}:} Left: Probability distribution of core radii. 
Right: Probability distribution of core line width. \\

\begin{figure}
\centerline{\epsfxsize=18cm \epsfbox{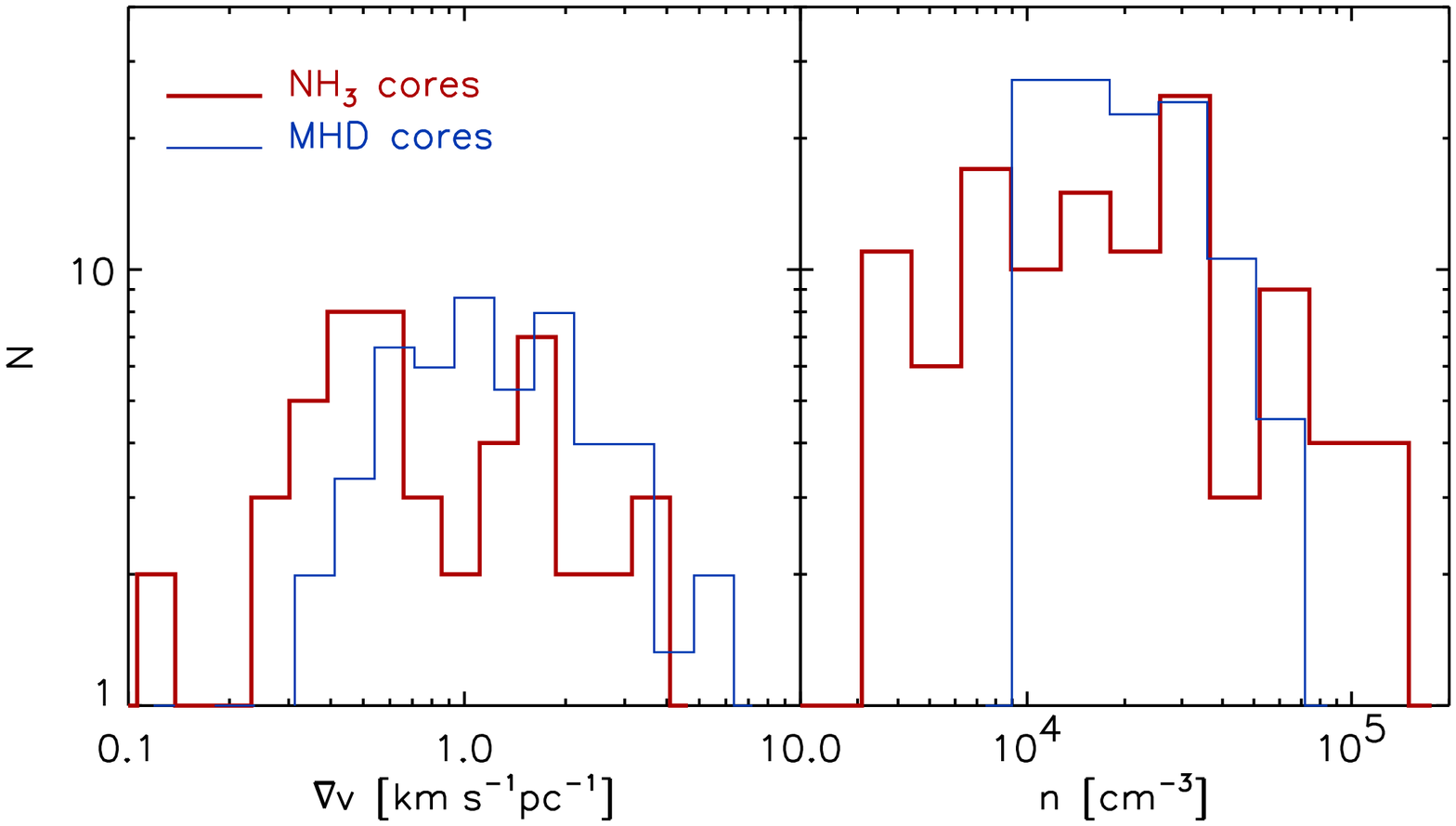}}
\caption[]{}
\label{fig3}
\end{figure}

{\bf Figure \ref{fig3}:} Left: Probability distribution of core velocity 
gradient. Right: Probability distribution of core density. \\

\begin{figure}
\centerline{\epsfxsize=18cm \epsfbox{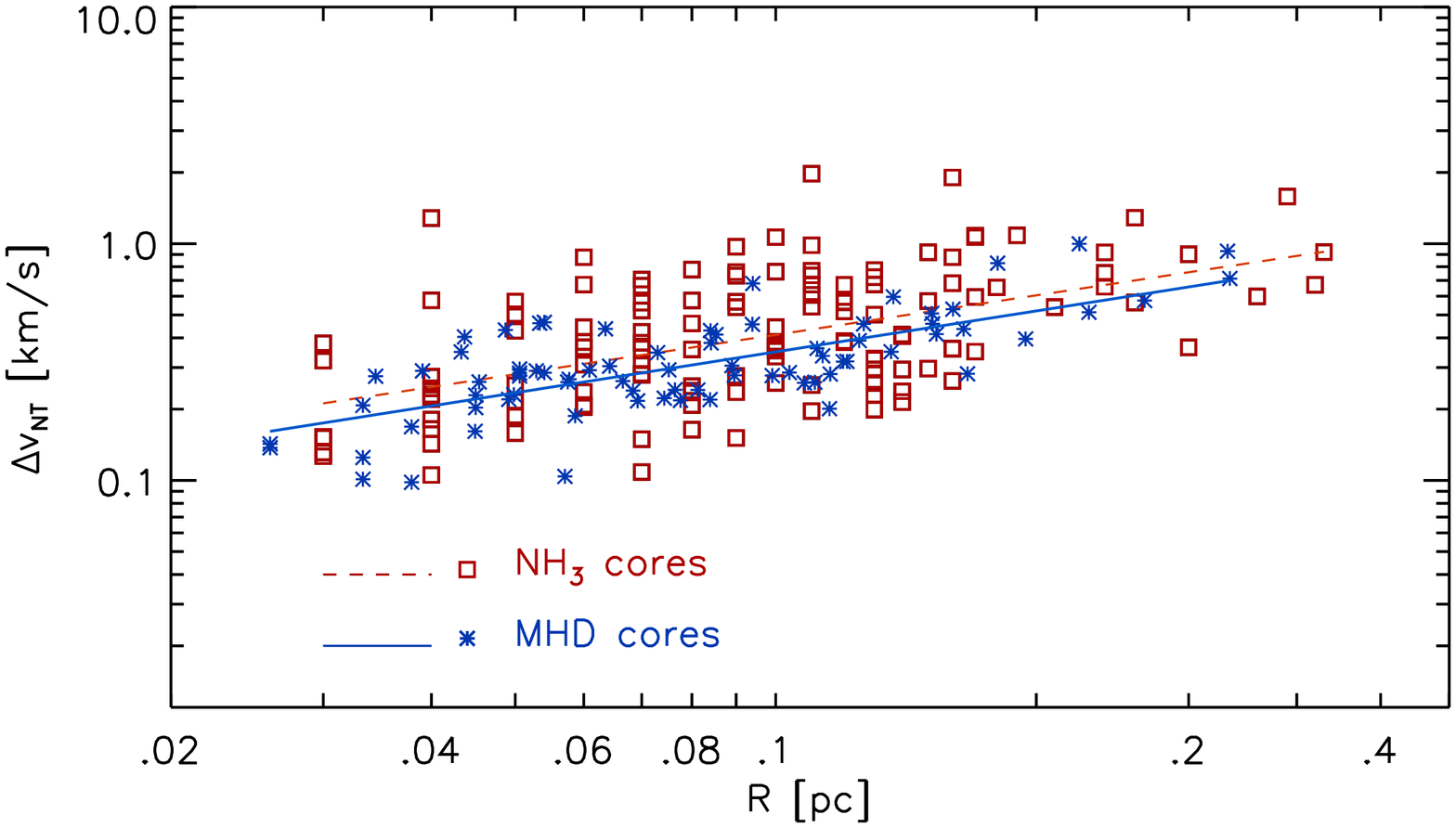}}
\caption[]{}
\label{fig4}
\end{figure}

{\bf Figure \ref{fig4}:} Non--thermal line width versus size for numerical 
cores (asterisks) and observed NH$_3$ cores (squares). \\

\begin{figure}
\centerline{\epsfxsize=18cm \epsfbox{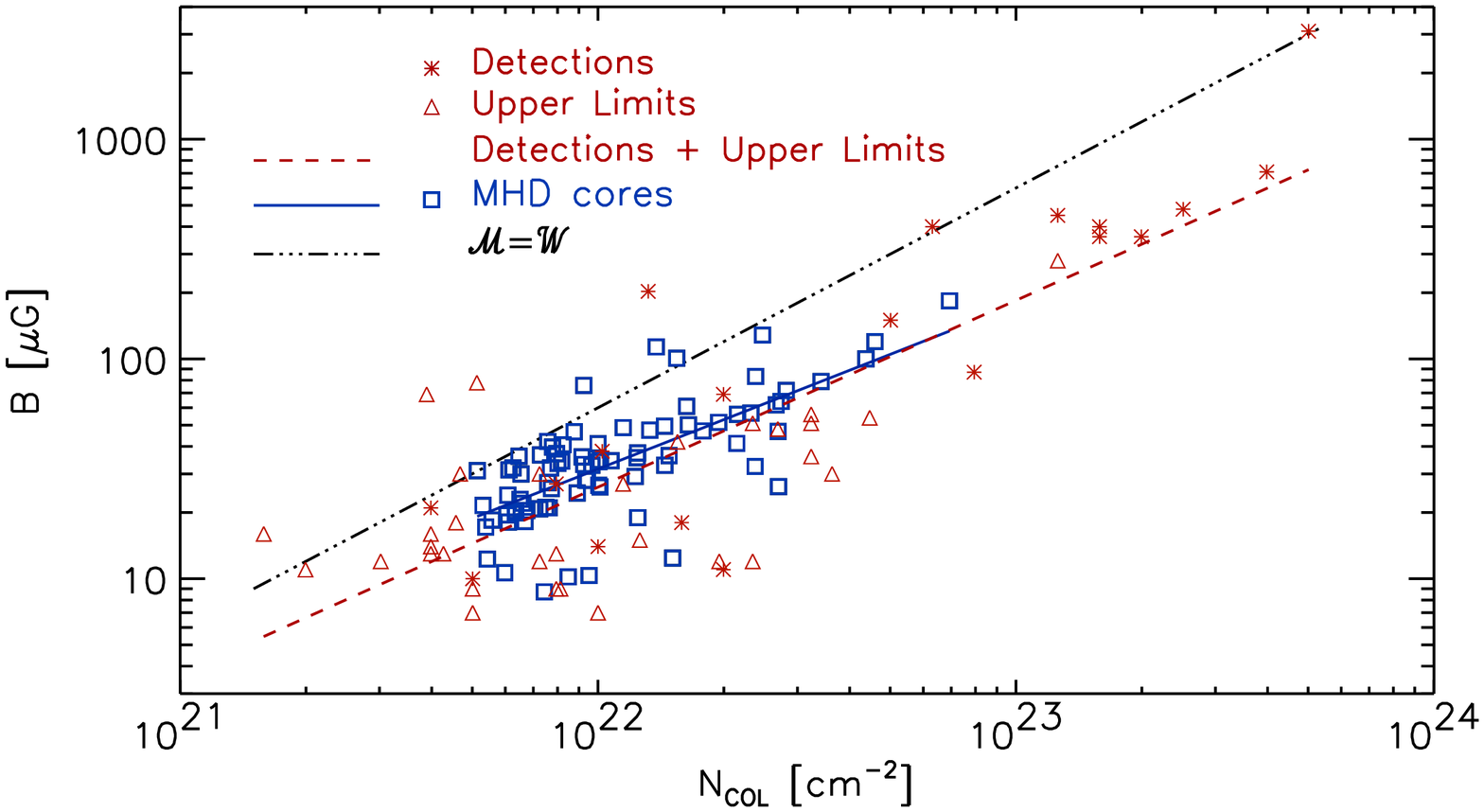}}
\caption[]{}
\label{fig5}
\end{figure}

{\bf Figure \ref{fig5}:} Magnetic field strength versus H$_2$ column density. 
Asterisks represent the Zeeman splitting measurements, and triangles
upper limits (19 detections and 31 upper limits). The dashed line is a 
least square fit to the Zeeman splitting measurements, where upper limits 
have been treated as if they were detections. The solid line is a least 
square fit to the MHD cores (squares). The dotted--dashed line marks the 
equality between magnetic and gravitational energies. \\

\clearpage
\begin{table}
\begin{tabular}{lccccc}
\hline
\hline
Cores  & $R$ [pc]      & $\Delta v$ [km/s] & $\nabla v$ [km/(s\,pc)] & $n$ [10$^4$~cm$^{-3}$] & $\Delta v$--$R$ slope \\ 
\hline
NH$_3$ & $0.11\pm0.07$ & $0.48\pm0.33$     & $1.42\pm1.39$           & $3.3\pm4.0$ & $0.56\pm0.22$  \\
MHD    & $0.09\pm0.06$ & $0.33\pm0.18$     & $1.84\pm1.56$           & $3.0\pm2.2$ & $0.57\pm0.15$  \\
\hline
\end{tabular}
\caption{}
\end{table}

\clearpage

\clearpage

\clearpage

\clearpage

\clearpage

\end{document}